# Mining and Analyzing the Future Works in Scientific Articles

Yue Hu and Xiaojun Wan

Institute of Computer Science and Technology, The MOE Key Laboratory of Computational Linguistics, Peking University, Beijing, China

{ayue.hu, wanxiaojun}@pku.edu.cn

**Abstract**

Future works in scientific articles are valuable for researchers and they can guide researchers to new research directions or ideas. In this paper, we mine the future works in scientific articles in order to 1) provide an insight for future work analysis and 2) facilitate researchers to search and browse future works in a research area. First, we study the problem of future work extraction and propose a regular expression based method to address the problem. Second, we define four different categories for the future works by observing the data and investigate the multi-class future work classification problem. Third, we apply the extraction method and the classification model to a paper dataset in the computer science field and conduct a further analysis of the future works. Finally, we design a prototype system to search and demonstrate the future works mined from the scientific papers. Our evaluation results show that our extraction method can get high precision and recall values and our classification model can also get good results and it outperforms several baseline models. Further analysis of the future work sentences also indicates interesting results.

## Introduction

The future work section is an important part of a scientific article. The authors usually discuss how to extend their current works, approaches or evaluations in the future work section. These future works often contain valuable research information and give the researchers hints of new research directions or ideas. Future works indicate current interesting research directions. For example, deep learning is very hot in the latest years and many recent works on deep learning follow the future works in previously published papers. With a future work search engine, a researcher may quickly discover the future work directions about his interested research topics. So it is meaningful to mine the future works in the scientific articles.

To the best of our knowledge, no previous works have investigated the future work mining problem in the fields of information retrieval, text mining and digital libraries, and we are the first to explore this problem. In our current study, this problem can be decomposed into two parts: future work extraction and future work classification. Based on the extraction and classification results, we can further analyze the future works and we believe that we can find out important research information through future work mining. A search system can be developed to search and rank the extracted future works, and the search results can be displayed in different categories according to the classification results. In all, it is very useful and also very challenging to mine the future works in scientific articles.

In this paper, we focus on the techniques of future work extraction and classification. For future work extraction, we employ a regular expression based method. By observing a dataset in the computer science field, we define four future work categories: "*problem*", "*method*", "*evaluation*" and "*other*". The details of the four future work categories will be discussed in the next sections. So the future work classification task turns out to be a multi-class classification problem. This task is very challenging because there is much ambiguity in sentences of future work sections. We develop several kinds of useful features for multi-class classification of the future works, including n-gram features and template features.

Experimental results on a manually labeled dataset show that the regular expression based future work extraction method achieves very high precision and recall values and thus it is unnecessary to use a machine learning algorithm for future work extraction. We manually annotate the extracted future work sentences and construct a benchmark dataset for future work classification. Evaluation results on the dataset show that our proposed classification method outperforms the baseline methods and it can achieve acceptable results. In addition, to get better classification precision of a specified category, we employ a strategy to remove the instances that are not reliable. In this way, we can ensure further analysis of the future woks is more credible and the search results of a specific category are more accurate.

Furthermore, we apply our extraction method to the whole paper dataset and extract all the future works. The multi-class classification model is then used to classify the future work sentences into the four categories. We analyze the mined future works and discover some interesting results. In the end, we design a system to search and rank the future works. The future work results can be displayed in different categories.

The remainder of this paper is organized as follows. In Section 2 we discuss the related work. We analyze the problem and introduce our corpus in Section 3. Section 4 describes the future work extraction algorithm and shows the evaluation results. The future work classification task and the evaluation results are introduced in the Section 5. Further analysis is presented in

Section 6 and system design is introduced in Section 7. We conclude our work and discuss our future work in Section 8.

## Related Work

Future work extraction and classification is a new area and there is no existing work on the field. However, there are many coarsely related research topics about information extraction and mining from scientific articles in the digital library, document recognition and text mining fields. Till now, machine learning methods like Support Vector Machines (SVM), Hidden Markov Model (HMM), Conditional random field (CRF), Markov Logic Networks (MLN) have been widely applied to various tasks of information extraction from research papers (Han *et al.*, 2003; Dalvi *et al.*, 2012; Peng and McCallum, 2006; Giles *et al.*, 1998; Seymore *et al.*, 1999; Huang and Wan, 2013]. Other research topics include paper summarization (Qazvinian and Radev, 2008; Abu-Jbara and Radev, 2011), survey generation (Mohammad *et al.*, 2009; Yeloglu *et al.*, 2011), literature search (Li *et al.*, 2010; Shahaf *et al.*, 2012), and paper recommendation (Sugiyama *et al.*, 2010).

Though future work section has not been investigated yet, other sections in scientific articles such as acknowledgement and related work sections have been explored. Early works on acknowledgment extraction were carried our manually. Cronin et al. (1993) investigated prominent journals of sociology for 10 years period and manually extracted acknowledgments. They also conducted similar studies on psychology and philosophy journals for a 100 years span in (Cronin *et al.*, 2003). Six categories of acknowledgement have been proposed: 1) moral support, 2) financial support, 3) editorial support, 4) presentational support, 5) instrumental/technical support, and 6) conceptual support or peer interactive communication

(PIC). PIC has been considered as the most important one for identifying intellectual debt (McCain, 1991). The acknowledgements are also investigated in other fields including history (Scrivener, 2009), information sciences (Cronin, 2001), and humanities (Cronin *et al.*, 1993).

Automatic acknowledgement extraction is discussed in (Councill *et al.*, 2005; Giles *et al.*, 2004). They regarded the problem of extracting acknowledgement from research articles as a specific case of document metadata extraction or information extraction from research papers. Regular expressions and machine learning algorithms are used for the acknowledgement extraction. They found that regular expressions work acceptably well for identifying acknowledgements within identifiable acknowledgement passages. This conclusion is consistent with the results of our future work extraction. Since a significant portion of acknowledgements is found in unlabeled passages, they adopt machine learning algorithms like Support Vector Machines (SVM) to identify these acknowledgements. Khabsa *et al.* (2012) proposed a system named AckSeer which is a repository and search engine for automatically extracted acknowledgements form digital libraries. It is a fully automated system that scans items in digital libraries including conference papers, journals, and books extracting acknowledgement sections and identifying acknowledged entities mentioned within.

Research on the related work sections mainly focuses on the summarization of related work. Hoang and Kan (2010) proposed a related work summarization system given the set of keywords arranged in a hierarchical fashion that describes the paper's topic. They used rule-based strategies to generate the related work section. Hu and Wan (2014) introduced a system called ARWG to generate the related work section for the original target paper. They used supervised learning and an optimization framework to deal with the task.

Wainer and Valle (2013) surveyed the topic of what happens to computer science research after it is published. They investigated the proportion of papers that continue, how and when the continuation takes place and whether any distinctions are found between the journal and conference populations by using self-citation practices. They found out 23% (journal) and 30% (conferences) of the papers published in CS have extensions. The result indicates that most research directions mentioned in the future work sections have not been investigated by their original authors and can potentially be adopted by other researchers.

## PROBLEM ANALYSIS AND CORPUS

**Problem Analysis**

Future work discussion is important in a scientific paper. Some authors introduce their future work in the future work section, while other authors often discuss their future works after concluding their current works in the conclusion section.

In our opinion, a future work sentence always contains unique information while the future works of one paper usually describe several research directions. To simplify our problem, we treat one future work sentence as the basic unit of the future works in one paper. We have the intuition that the future works can be classified into different categories. After observing the data in the computer science field, we classify the future work sentences into the following four categories according to the main contents of the sentences:

1) The "Problem" Category

This category includes future work sentences described as below:

- Extend the current work: e.g., "*In future work, we would like to extend our work on embedding SMT in virtual gameplay to larger and more diverse datasets, and involve human feedback in the response-based learning loop.*"

- Apply the proposed approach to other tasks or datasets: "*Also, we will apply our model to additional opinion analysis tasks such as fine-grained opinion summarization and relation extraction.*"

- Introduce a new research problem: "*For future work we intend to investigate how syntax-based information can be used to introduce more semantic structure into the graph.*"

2) The "Method" Category

This category includes future work sentences described as below:

- Extend the proposed approach: "*In the future, we plan to extend our model to capture both grammatical and lexical cohesion in document-level machine translation.*"

- Introduce a new approach: "*We would also like to employ lexicalized models that should help in situations in which the pos tags are too coarse.*"

3) The "Evaluation" Category

This category includes future work sentences described as below:

- Evaluate on the new datasets or sources: "*For future work we intend to evaluate our method on other datasets and domains, varying in level of language complexity and correctness.*"

- Employ other evaluation metrics: "*Concerning qualitative evaluation, we will try to apply evaluation metrics that are able to capture content and coherence aspects of summaries, such as more complex content similarity or readability measures.*"

4) The "Other" Category

The other future work sentences that do not belong to the above three categories are included in the other category. For example, *"Besides, we plan to collect more Chinese-Japanese patent corpus as the currently available corpus size is still too small."* and *"Our future work will attempt to develop an open-domain opinion web search engine."*

**Corpus and Preprocessing**

We build a corpus that contains academic papers that are downloaded from the ACL Anthology [1]. The ACL Anthology includes all papers published by ACL and related organizations as well as the Computational Linguistics journal over a period of four decades. It currently hosts over 33,000 papers from major conferences such as ACL, EMNLP, COLING in the fields of computational linguistics and natural language processing (NLP). Though NLP is a subfield of the computer science field, the corpus in the NLP field has been widely adopted for many text mining tasks and the researches on this corpus are very representative and without loss of generality.

We select the papers in four different domains: "*parse*", "*machine translation*", "*emotion analysis*" and "*summarization*" from the whole paper set of the ACL Anthology. We match the title of the paper with some keywords to identify whether it belongs to these four domains and which domain it belongs to. For example, if the title of one paper contains keywords like "*emotion*", "*opinion*" or "*sentiment*", we add this paper into the emotion analysis domain. The numbers of papers we select for the four domains are shown in Table 1. The total papers in all four domains constitute our raw corpus.

---

[1] http://aclweb.org/anthology/

The papers are all in PDF format. We extract their texts by using PDFlib[2] and detect their physical structures of paragraphs, subsections and sections by using ParsCit[3]. If the paper has the future work section, we directly extract it. If not, we extract the conclusion section, as many authors discuss their future work in the conclusion section.

Table 1

*The paper number selected for each domain*

| Domain | Number |
|---|---|
| Parse | 1254 |
| Machine translation | 1476 |
| Emotion analysis | 572 |
| Summarization | 406 |

**FUTURE WORK EXTRACTION**

In this section we describe the method employed for extracting future work sentences. As mentioned above, we have already extracted the future work section or conclusion section from each paper. The conclusion sections usually contain sentences that conclude the authors' own work. In addition, there exist sentences that are not related to the future works even in the future work sections. We want to identify the future work sentences that contain valuable information and discard useless sentences from the given sections.

The task of extracting future work sentences can be viewed as a special case of automatic document metadata extraction. Several approaches have been proposed for automatic document metadata extraction, including regular expressions, rule-based parsers and machine learning algorithms. Regular expressions and rule-based parsers can be easily implemented and perform well if data are well behaved. We have found out regular expressions and rules work well for identifying the future work sentences from the future work or conclusion sections.

---

[2] http://www.pdflib.com/

[3] http://aye.comp.nus.edu.sg/parsCit/

Algorithm 1 shows the pseudo code of our future work sentence extraction process. We define two different tiers of regular expressions, and the first tier is stricter than the second tier. The sentence that match the regular expressions in the first tier is considered to have a very high probability to be a future work sentence and it is directly included into the future work sentence set. Once one sentence that matches the regular expressions in the first tier is found, the sentence that matches the regular expressions in the second tier can also be considered to be a future work sentence. The reason why we define these two tiers is that the regular expressions in the second tier can also match the sentences which are related to the conclusion. The authors usually conclude their current works and then discuss their future works separately in the conclusion sections. So if one future work sentence is found, we consider the authors have started to discuss their future works and the sentences after this future work sentence are probably related to the future works if it contains a regular expression in the second tier. The examples of the two regular expression tiers we use are shown in Table 2. Another issue is that we may extract some valueless future work sentences. For example, "*There are many directions of future work to pursue here.*" We remove these sentences from the extracted future work sentence set by using some rules.

To evaluate the effectiveness of our extraction algorithm we need an annotated dataset. To the best of our knowledge, there is no existing gold standard dataset for future work extraction. So we construct a dataset manually. Our dataset contains about 400 papers selected from our raw corpus. Each domain contains about 100 papers. We annotate the future work sentences in these papers. The number of the annotated future work sentences is 724 after removing near-duplicate sentences. We also implement a baseline method that treats the second tier of regular expressions as the first tier.

Table 3 shows the comparison results of our future work extraction algorithm and the baseline method on the annotated dataset. Our regular expression based extraction algorithm outperforms the baseline method and yields precision, recall and f-measure values at 85.6%, 94.7% and 89.9%, respectively. It proves that our assumption of future works is reasonable. Some sentences are extracted by mistake and the main reason is that some authors mix the conclusion and future work sentences in their conclusion sections. In this case, some conclusion sentences that match the second tier of regular expressions may be extracted by mistake. Overall, we can find out that the extraction algorithm based on regular expressions outperforms the baseline method and performs well on our data. It is not necessary to employ a machine learning method such as SVM.

Table 4 shows the performance of the future work extraction algorithm on each domain. We can see that our extraction algorithm can get high performance values on all the four domains. The precision of the algorithm in the emotion analysis domain is a little lower than that in the other domains. The possible reason is that some authors in the emotion analysis domain like to discuss their current work and future work in a mixed way.

**ALGORITHM 1.** Future work extraction algorithm

**Input:**
A future work or conclusion section *F* that contains a sentence set *S*;
*TR1*: the regular expressions in the first tier;
*TR2*: the regular expressions in the second tier;
**Output:**
The future work sentence set *FS*;
**Procedure:**
*FWFound = FALSE*;
**while** not at end of *F* **do**
    *read sentence s from S;*
    **if** *s matches regexes in TR1* **then**
  Add *s* to *FS*;
  *FWFound = TRUE;*
    **else**
  **if** *FWFound == TRUE*
        **if** *s matches regexes in TR2* **then**
           Add *s* to *FS*;
        **end**
    **end**
      **end**
**end**

Table 2

*The examples of regular expressions*

| Tier | Examples |
|---|---|
| 1 | *we…will, we…plan, we…want, we…wish, future, to extend and other ten regular expressions* |
| 2 | *we…can, we…believe, we…could, is to, further, suggest, promising, potential and other eleven regular expressions* |

Table 3

*Comparison results for future work extraction*

| Method | Precision | Recall | F-measure |
|---|---|---|---|
| Baseline | 61.7% | 95.6% | 75.0% |
| Our method | 85.6% | 94.7% | 89.9% |

Table 4

*Performance of our future work extraction algorithm on each domain*

| Domain | Precision | Recall | F-measure |
|---|---|---|---|
| Parse | 85.6% | 93.6% | 89.4% |
| Machine translation | 86.5% | 94.3% | 90.2% |
| Emotion analysis | 81.6% | 95.2% | 87.9% |
| Summarization | 90.2% | 95.6% | 92.8% |

## FUTURE WORK CLASSIFICATION

In this section we will investigate the future work classification task. As mentioned in the above sections, we have extracted the future work sentences and we have defined four different future work categories: "*problem*", "*method*", "*evaluation*" and "*other*". In order to further analyze the future works and display the search results by category, the future work classification task is necessary.

**Data Annotation**

To train and evaluate the classification methods we also need an annotated dataset. There is no existing gold standard dataset for future work classification and so we build our classification dataset based on the dataset annotated for future work extraction. We annotate each future work sentence in the context of the corresponding paper and assign each sentence with a tag that indicates its category.

Table 5 shows the distribution of annotated sentences in our classification dataset. We can see that most future work sentences belong to the first three categories. The problem and method categories contain about 75 percent of the total future work sentences. This result proves that our definition of the future work sentence categories is reasonable.

Table 5

*The distribution of annotated sentences*

| Category | Number | Percentage |
|---|---|---|
| Problem | 273 | 37.7% |
| Method | 269 | 37.2% |
| Evaluation | 129 | 17.8% |
| Other | 53 | 7.3% |

**Classification**

The future work classification can be treated as a multi-class classification problem because we have defined four different categories. In this subsection, we first briefly introduce the multi-class classification problem and then describe features we use. Lastly, we present the strategy to improve the classification precisions for the first three categories.

**Multi-class classification.**

Classification tasks aim to assign a predefined class label to each instance. In cases when the number of the class labels equals two, the classification problems are referred as binary classification problems. But many real-word problems have more than two classes to deal with. These classification problems are regarded as multi-class classification.

There have been many studies on multi-class classification. In general, multi-class classification methods can be roughly partitioned into two groups. The first group of methods can be naturally extended to handle multi-class classification, such as regression, nearest neighborhoods (Hastie *et al*., 2005) and decision trees including C4.5 (Quinlan, 1993) and CART (Olshen and Stone, 1984). The second group of methods involves reduction of multi-class classification problems to binary ones. There are several reduction techniques that can be used such as one-versus-the-rest method (Bottou *et al*., 1994; Scholkopf *et al*., 2002), pairwise comparison (Friedman, 1996; Hastie and Tibshirani, 1998; Kreßel, 1999), direct graph traversal (Platt *et al*., 1999), error-correcting output coding (Allwein *et al*., 2001; Dietterich, 1995) and multi-class objective functions (Franc and Hlaváč, 2002). In practice, the reduction method used depends on the underlying problem.

In this study, we adopt the popular pairwise comparison method to complete the reduction. In pairwise comparison, a classifier is trained for each possible pair of classes. So for K classes, $(K - 1)K/2$ binary classifiers are needed to be trained. Given a new instance, the multi-class classification is then executed by testing all $(K − 1)K/2$ individual classifiers and assigning an instance to the class which gets the highest number of votes. The class number of our future work classification is not large. The number of the individual classifiers that are needed to be trained is also not large. The individual classifiers are usually easier to be learned, and we think the pairwise comparison method is a good choice for our classification problem. In particular, we use LIBSVM (Chang and Lin, 2011) to realize our multi-class classification.

**Features.**

We develop four groups of features for the future work classification task: unigram features, bigram features, manually defined template features, and automatic extracted template features. The features are described as follows:

1) **Unigram features**: The unigram features capture clues between the lines. They are usually effective in most of situations. We remove the general stop words besides the specific stop words for the future works such as "*future*". We select only a proportion of the unigrams that occur frequently in the dataset. The unigram feature set contains 496 unique unigrams and we use a binary feature value to indicate whether the sentence contains the corresponding unigram.

2) **Bigram features**: Similar to the unigram features, we retain only a subset of the bigrams that appear in the dataset frequently. The bigrams that contain only the stop words are also dropped. In addition, some special bigrams such as "*future work*" are removed, for they are not useful for our classification task. The bigram feature set contains 423 unique bigrams and we also use a binary feature value to indicate whether the sentence contains the corresponding bigram.

3) **Manually defined template features**: We define a few templates manually and they have three types: words, phrases and regular expressions. We define different template set for each type and each category. Because there is no regular expression template set for the "*evaluation*" category, we actually get eleven different template sets. Table 6 shows some examples for each template set. For each instance (i.e., future work sentence), we count the matching number of the templates in each template set and set the number as the corresponding feature weight. So the number of features we add into the whole feature set is eleven, which is the same as the number of the manually defined template feature sets.

4) **Automatic extracted template features**: Though we have defined templates manually, these templates can achieve good precision but low recall since it is not trivial to discover all the templates manually. So we also extract a number of templates from the annotated dataset automatically. Different from the manually defined templates, these automatic templates are directly added into the whole feature set as the unigram and bigram features. Binary feature values are also employed. The number of these features is 565. We first tokenize the future work sentences, stem the tokens and treat each sentence as a word sequence. We apply the PrefixSpan (Prefix-projected Sequential Pattern Ming) algorithm (Pei *et al*., 2001) which is used to find frequent sequential patterns in sequential data. PrefixSpan explores prefix-projection in sequential pattern mining. Its general idea is to examine only the prefix subsequences and project only their corresponding postfix subsequences into projected databases. Sequential patterns are grown by exploring only local frequent patterns. By setting the parameters including the minimum length and the minimum support (i.e. occurrence frequency), we can get sequential patterns whose length is no less than the minimum length and whose occurrence count is no less than the minimum support. We set the minimum length to two and set the minimum support to two. We remove the patterns which only contain stop words because we think they are useless for our classification task. In order to get useful templates for each category, we employ the PrefixSpan method to each future work sentence set with the same category. After merging all the template sets we get for the different categories, we get our final automatic extracted templates. Algorithm 2 shows the outline of our extraction method.

**ALGORITHM 2.**   Automatic Template Extraction
--------
**Input:**
Future work sentence set *FS*;
Parameters: minimum length, minimum support
**Output:**
Template set;
**Procedure:**
Remove sentences with less than 5 words in length;
Tokenize and stem sentences into word sequences
*total_template_list = { };*
**for** sentence set *S* of each category in *FS* **do**
    *template_list = PrefixSpan(S, minimum length, minimum support);*
    **for** *template* in *template_list* do
        **if** (*template* are not all stopwords) **then**
            add *template* into *total_template_list*;
   **end**
    **end**
**end**
Remove the duplicate templates in *total_template_list;*
**return** *total_template_list*;

Table 6

*The examples of templates defined*

| Class + Type | Examples |
|---|---|
| Problem + Word | *apply, capture, task, direction, domain, topic, area, issue, contribution, ...* |
| Problem + Phrase | *aim at, further research, interesting to investigate, find out, later work, ...* |
| Problem + Regex | *(other/another).*?(task/domain/language/problem/work)* <br> *(extend/employ/apply/explore).*?(our/this).*?(work/problem)* |
| Method + Word | *adapt, incorporate, integrate, combine, enhance, joint, adjust, refine, ...* |
| Method + Phrase | *our model, our method, make use of, sophisticated feature, joint model,...* |
| Method + Regex | *(investigate/add/include/weight/introduce/explore/refine).*?feature* <br> *(explore/improve/extend/try/refine/exploit).*?(appro* |

|  | *ach/method/model/algorithm)* |
| --- | --- |
| Evaluation + Word | *evaluate, verify, identify, performance, affect, examine, feature, experiment,…* |
| Evaluation + Phrase | *failure analysis, more language, different parallel, error analysis, error type,…* |
| Other + Word | *enrich, creation, application, version, available, code, release, …* |
| Other + Phrase | *source code* |
| Other + Regex | *(create/gather/publish/release/complete/provide/collect).*?(corpus/set/corpra/software/system) will be.*?(available/release)* |

**Classification precision improvement.**

With the annotated dataset and the feature set we propose, we use the LIBSVM toolkit for model training and testing. Since the "*other*" category has many mixed future work sentences, we think the first three categories which are "*problem*", "*method*" and "*evaluation*" are more important than the "*other*" category. As we intend to further analyze the future works and display the search results in our system by category, classification precision is more important than recall. Better classification precision can lead to more reliable analysis results and better search results. So we intend to improve the classification precisions of the first three categories.

As mentioned above, the multi-class classification method we adopt tests all $(K − 1) K /2$ individual classifiers (K is the number of the classes) and assigns the instance to the class which gets the highest number of votes (i.e. the highest possibility) when given a new instance. For the instances that are difficult to classify, the corresponding values of the highest possibility are often lower than that for the instances that are easy to classify.

So, in order to improve the classification precisions of the first three categories, we post-process the classification results. For instances whose highest possibilities exported by the LIBSVM toolkit are lower than a predefined threshold, we move the instances to the "*other*" category. In this way, we can ensure that the future work sentences classified into the first categories are more accurate and they are suitable for further analysis and demonstration.

**Experiment and Analysis**

   **Experiment setup.**

We use the annotated future work sentence set for our experiments. As mentioned earlier, the dataset contains 724 sentences and the percentages of the "*problem*", "*method*", "*evaluation*" and "*other*" categories are 37.7%, 37.2%, 17.8% and 7.3%, respectively. Five-fold cross-validation is performed on the dataset and the reported performance values are average values across the five folds.

In the experiments, we compare our proposed approach with the following two baseline methods to evaluate the effectiveness of our approach.

   1) SVM with unigram and bigram features (SVM1): Here we use only the unigram and bigram features as the baseline feature set. These features are the most widely-used features in traditional text classification tasks.

   2) SVM with the unigram, bigram and manually defined template features (SVM2): Besides the unigram and bigram features used in SVM1, we include the manually defined template features into the feature set.

In addition, in order to improve the classification precision, we use a threshold to control the lower bound of the possibility of the future work sentences that can be classified into the first three categories. Here we try different values of the threshold to observe its impact on the classification results.

**Performance measure.**

In general, precision, recall and F-measure are used for performance measurement. For multi-class classification problem, we usually use micro average and macro average for evaluation. The micro average and macro average on precision, recall and F-measure are computed as follows:

$$Micro\_Precision = \frac{\sum_i \#system\_correct(class = i)}{\sum_i \#system\_proposed(class = i)}$$

$$Micro\_Recall = \frac{\sum_i \#system\_correct(class = i)}{\sum_i \#gold(class = i)}$$

$$Micro\_F - measure = \frac{2 * Micro\_Precision * Micro\_Recall}{Micro\_Precision + Micro\_Recall}$$

$$Macro\_Precision = \frac{1}{4} \sum_i \frac{\#system\_correct(class = i)}{\#system\_proposed(class = i)}$$

$$Macro\_Recall = \frac{1}{4} \sum_i \frac{\#system\_correct(class = i)}{\#gold(class = i)}$$

$$Macro\_F - measure = \frac{2 * Macro\_Precision * Macro\_Recall}{Macro\_Precision + Macro\_Recall}$$

*#system_proposed(class=i)* is the number of future work sentences classified into the *i*-th class. *#system_correct(class=i)* is the number of future work sentences correctly classified into the *i*-th class. *#gold(class=i)* is the number of future work sentences manually annotated in the *i*-th class. *i* indicates one of the four future work categories.

Table 7

*Comparison results for future work classification*

| Method | Micro average | | | Macro average | | |
|---|---|---|---|---|---|---|
| | Precision | Recall | F-measure | Precision | Recall | F-measure |
| SVM1 | 0.6623 | 0.6623 | 0.6623 | 0.6560 | 0.5742 | 0.6123 |
| SVM2 | 0.7143 | 0.7143 | 0.7143 | 0.7324 | 0.6656 | 0.6971 |
| Our Method | 0.7493 | 0.7493 | 0.7493 | 0.7458 | 0.6755 | 0.7085 |

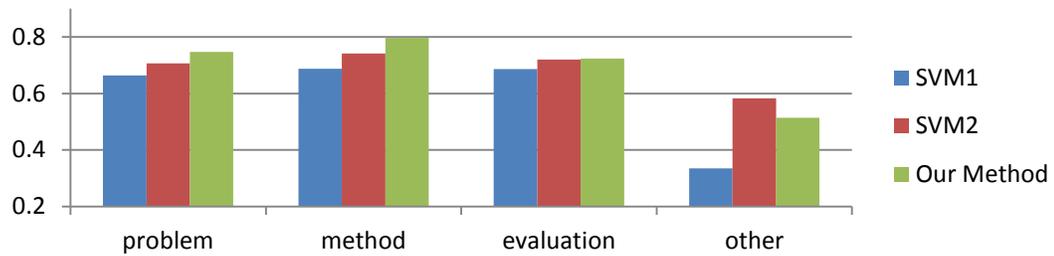

*Figure 1*. F-measure on four future work categories for different approaches

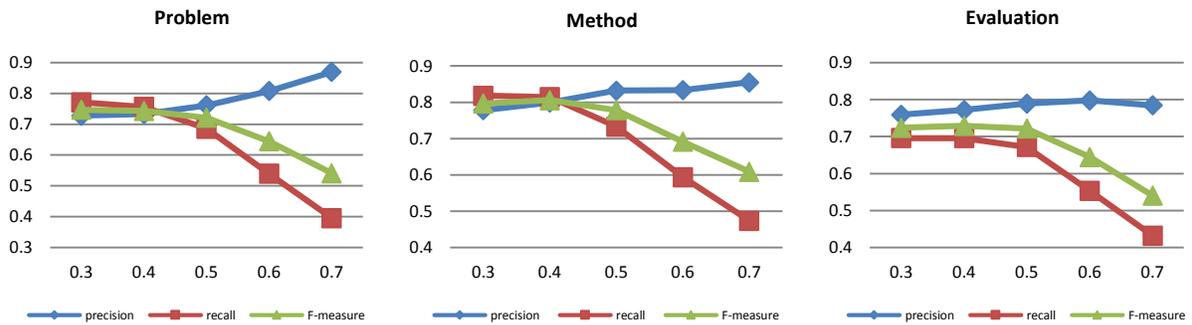

*Figure 2*. Impact of the threshold on the classification results

**Results and analysis.**

Table 7 shows the comparison results of our method and the baseline methods with default settings (i.e., the classification precision improvement is not applied). Our method outperforms all baseline methods over both macro average F-measure and micro average F-measure. The micro average precision and recall values are the same in our case because we always assign a category to an instance and then the total number of future work sentences classified is equal to that of future work sentences manually annotated. Comparing the results of SVM1 and SVM2, we can find out that SVM2 performs better than SVM1, which proves the manually defined template features are useful for the classification task. We can also get the conclusion that the automatic template features are also effective for the classification task, because our method can get better results than SVM2. With all the features, our method can get the best classification results.

Figure 1 shows the F-measure scores of our method and baselines methods on the four different categories. We can see that our method can achieve better performances on "*problem*", "*method*" and "*evaluation*" categories. Since the future work sentences in the "*other*" category are highly mixed and significantly less than the sentences in the first three categories, it is not trivial to classify them correctly. Our method performs not very well on the "*other*" category. This is also the reason why the macro average results are a little worse than the micro average results. Overall, our method outperforms the baseline methods on the three important categories.

Figure 2 shows the impact of the threshold that controls the lower bound of the possibility on the classification results. When the threshold is less than 0.3, the results almost remain unchanged. When the threshold is bigger than 0.7, the classification results are too bad. So we show the results when the threshold is set between 0.3 and 0.7. We can see that when the threshold is set to 0.5, the precisions are significantly better and the F-measure scores can nearly remain unchanged. The precisions of the "*problem*", "*method*" and "*evaluation*" categories raise from 0.728 to 0.761, from 0.778 to 0.832 and from 0.759 to 0.789, respectively. The F-measure scores of the "*problem*", "*method*" and "*evaluation*" categories change from 0.747 to 0.721, from 0.797 to 0.778 and from 0.724 to 0.722, respectively. So 0.5 is a reasonable value for the threshold.

## FUTURE WORK ANALYSIS

Table 8

*The future work sentence stats for each domain*

| Domain | Paper number | Future work sentence number | Average sentence number per paper |
| --- | --- | --- | --- |
| Parser | 705 | 2011 | 2.85 |
| Machine translation | 878 | 2448 | 2.79 |
| Emotion analysis | 374 | 946 | 2.52 |
| Summarization | 258 | 802 | 3.1 |

Table 9

*The future work category distribution for each domain*

| Domain | Category | | | |
| --- | --- | --- | --- | --- |
| | Problem | Method | Evaluation | Other |
| Parse | 41.4% | 43.0% | 11.3% | 4.3% |
| Machine translation | 42.7% | 39.9% | 14.9% | 2.5% |
| Emotion analysis | 41.6% | 39.8% | 17.3% | 1.2% |
| Summarization | 43.3% | 39.7% | 13.1% | 3.7% |

We employ the future work extraction algorithm to our whole raw corpus and extract future work sentences. Then we train the multi-class classification model on the whole annotated dataset and apply the model to the extracted future work sentences. The strategy to improve the classification precision of the first three categories is also used. Table 8 shows the stats of the extracted future work sentences on each domain. We can see that each paper has about three future work sentences in average. Table 9 shows the distribution of the future work categories on each domain. We can find out that this distribution is generally consistent with the distribution of the annotated data. The differences are mainly because our classification model performs not very well on the "*other*" category, which leads some sentences of the "*other*" category to be classified into the wrong categories.

To further analyze the future works, we try to extract the academic concepts in the future works. It is not trivial to accurately extract academic concepts and it is beyond the scope of this paper. Instead, we simplify the goal and attempt to extract the important noun phrases in the future works. We think these noun phrases can include most academic concepts. To extract the important noun phrases, phrase chunking implemented by the OpenNLP[4] library is applied to the sentences. All the phrases are stemmed. We then remove the prefix stop words in each phrase. The phrases that only contain stop words are discarded. The phrases that contain only one word which can be found in the WordNet are also dropped which are always common phrases. The remaining phrases are counted and used.

---

[4] http://opennlp.apache.org/

*Figure 3.* Word cloud for "*emotion analysis*"

*Figure 4.* Word cloud for "*parse*"

*Figure 5.* Word cloud for "machine translation"

*Figure 6.* Word cloud for "summarization"

Figure 7: Word cloud for "*parse*" in "*problem*" category

*Figure 8.* Word cloud for "*parse*" in "*method*" category

Figure 3 through Figure 6 shows the word cloud of each domain. We can find useful information in the word clouds. For example, we can discover phrases such as "*Hashtag*" and "*Twitter Data*" which are popular in the domain of emotion analysis in the word cloud of emotion analysis. It implies that the future researches in the emotion analysis domain will focus more on emotion analysis on social media content (e.g. Twitter). We can find the phrase "*machine translation*" in both the "*parse*" domain and "*summarization*" domain which indicates the strong relationship between the "*machine translation*" domain and these two domains. Figure 7 and Figure 8 shows the word cloud of the "*parse*" domain in the "*problem*" and "*method*" categories, respectively. In the word cloud of the "*problem*" category we can see the phrases such as "*machine translation*", "*statistical parser*", "*supervised parsing*", "*shallow parser*" which are related to the "*problem*" category, while in the word cloud of the "method" category, phrases such as "*elementary tree*", "*morphological feature*", "*probabilistic model*" can be found.

## SYSTEM DESIGN

In the end, we design a prototype system to search and demonstrate the future works. The interface of our system is shown in Figure 9. With the system, users can directly browse the future works in a specific domain by selecting a domain name in the domain list which contains "*Emotion analysis*", "*Summarization*", "*Parse*", "*Machine translation*", and other domains. Users can also type a query and the matched future works will be returned.

Moreover, users can also select a ranking method to rank the returned results, and the future work sentences can be ranked according to the sentence's PageRank value, the corresponding paper' publication date, and the corresponding paper's citation number. In our opinion, the future work sentences which have higher PageRank scores are more important and the more recent future works are also more useful since they can capture the latest research directions. In addition, the future works of the more remarkable papers are considered to be more valuable and reliable. We use the citation number to measure the importance of a paper.

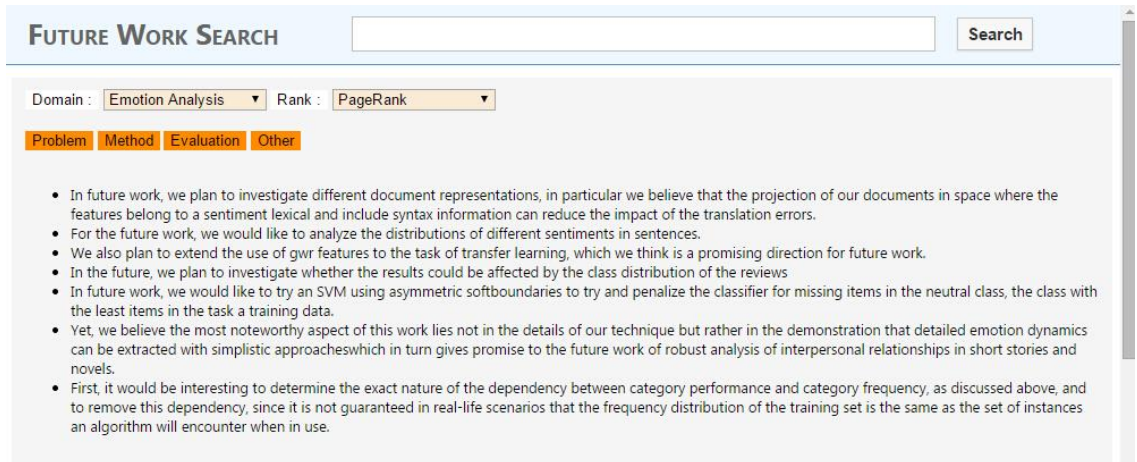

*Figure 9.* The interface of our system

Lastly, based on the classification results, the search results can be displayed in different categories. User can switch the search results by clicking the category tags. This kind of category-based future work display will help a user to quickly browse the search results and find the most import one he is interested in.

## CONCLUSION AND FUTURE WORK

In this paper, we mine and analyze the future works in the scientific articles. We first explore the problem of future work extraction. After defining four different categories of the future works, we investigate the multi-class classification problem of the future works. Then we further analyze the future works on the whole dataset. At last a prototype system is designed to search and demonstrate the future works. Evaluation results show that our future extraction method can get high precision and recall values and our classification model outperforms the baseline methods. Further analysis indicates that the future works can provide useful research information.

In the future work, we will polish our future work search system and release the system to the public. We will also find a way to better measure the importance of each future work by considering many factors, such as the contents, the venue information, the author information and the paper information.